\documentclass[onecolumn,conference]{IEEEtran}
\IEEEoverridecommandlockouts
\usepackage{cite}
\usepackage{amsmath,amssymb,amsfonts}
\usepackage{algorithmic}
\usepackage{graphicx}
\usepackage{textcomp}
\usepackage{tabularx}

\usepackage{xcolor}
\usepackage{graphicx}
\usepackage{cite}
\usepackage{picinpar}
\usepackage{amsmath}
\usepackage{url}
\usepackage{flushend}
\usepackage[utf8]{inputenc}
\usepackage{colortbl}
\usepackage{soul}
\usepackage{multirow}
\usepackage{makecell}
\usepackage{pifont}
\usepackage{xcolor}
\usepackage{alltt}
\usepackage[hidelinks]{hyperref}
\usepackage{enumerate}
\usepackage{siunitx}
\usepackage{breakurl}
\usepackage{epstopdf}
\usepackage{pbox}
\usepackage{amsfonts}
\usepackage{mathtools}
\usepackage{commath}
\usepackage{booktabs}
\usepackage{caption}
\usepackage{subcaption}
\usepackage{bm}
\usepackage{mathrsfs}
\usepackage{amsthm}
\theoremstyle{remark}
\newtheorem{assumption}{Assumption}

\usepackage[linesnumbered,ruled,vlined]{algorithm2e}

\newcommand*{\derivativewi}{\dot{\omega}_i^{\delta}}
\newcommand*{\wi}{\omega_i^{\delta}}
\newcommand*{\inertiaconstanti}{H_i}

\newcommand*{\inertiaconstant}[1]{H_{#1}}

\newcommand*{\pflow}[2]{P_{#1 \to #2}}
\newcommand*{\pielec}{P_i^e}

\newcommand*{\pimech}{P_i^m}
\newcommand*{\ibrprimarygain}{R_i^{\text{IBR}}}
\newcommand*{\derivativepimech}{\dot{P}_i^m}

\newcommand*{\communicationdelay}{T^d}

\newcommand*{\pifcrn}{P_i^{\text{FCR}}}
\newcommand*{\pifcrncapacity}{P_i^{\text{FCR}_{\text{max}}}}

\newcommand*{\gainfcrn}{K_{\text{N}}}

\newcommand*{\hatwi}{\hat{\omega}_i^{\delta}}
\newcommand*{\rangefcrn}{\omega_{\text{N}}}
\newcommand*{\unmeasuredpoweri}{P^{?}_i}
\newcommand*{\unmeasuredpower}[1]{P^{?}_{#1}}

\newcommand*{\derivativeunmeasuredpoweri}{\dot{P}^{?}_i}
\newcommand*{\hatunmeasuredpoweri}{\hat{P}^{?}_i}

\newcommand{\pirenewable}{P^{RES}_i}

\newcommand*{\actualcontrolsignal}{u^{m}_i}
\newcommand*{\derivativeactualcontrolsignal}{\dot{u}^{m}_i}
\newcommand*{\idealcontrolsignal}{u^{\star}_i}

\newcommand*{\distributedcontrolsignali}{u^{\Delta}_i}
\newcommand*{\derivativedistributedcontrolsignali}{\dot{u}^{\Delta}_i}
\newcommand*{\neighbourhoodi}{\mathcal{N}_i}
\newcommand*{\governortimeconstant}{T_i^g}

\newcommand*{\droopcoefficientsg}{R_i^{sg}}
\newcommand*{\reactancei}{X_{i,j}}
\newcommand*{\reactance}[2]{X_{#1,#2}}

\newcommand*{\powerimbalancei}{\delta_i^P}
\newcommand*{\powerimbalancedisti}{\delta_i^D}
\newcommand*{\hatpowerimbalancei}{\hat{\delta}_i^P}
\newcommand*{\node}[1]{\mathcal{V}_{#1}}
\newcommand*{\edge}[2]{\mathcal{E}}
 
\newcommand*{\timeconstantunmeasuredpower}{T_i^{?}} 
\newcommand*{\ibrtimeconstant}{T_i^{\text{BESS}}} 
\newcommand*{\piibrmin}{P_i^{\text{BESS}_{\text{min}}}}
\newcommand*{\piibrmax}{P_i^{\text{BESS}_{\text{max}}}}

\newcommand*{\actualpifcr}{P_i^{FCRm}}

\newcommand*{\scalingfactordistributed}[1]{K^{sf}_{#1}}

\newcommand*{\distributedcontrolgaini}{\alpha_i}
\newcommand*{\distributedcontrolsignalj}{u^{\Delta}_j}

\newcommand*{\communicationweightj}{w_{i,j}}

\def\BibTeX{{\rm B\kern-.05em{\sc i\kern-.025em b}\kern-.08em
    T\kern-.1667em\lower.7ex\hbox{E}\kern-.125emX}}

\begin{document}

\title{Coordinated Fast Frequency Regulation in Dynamic Virtual Power Plants via Disturbance Estimation\\ 
\thanks{Saif Ahmad and Seifeddine BEN ELGHALI are with Aix Marseille Univ, CNRS, LIS, Marseille, France.
Hafiz Ahmed is with Nuclear AMRC, Coventry, United Kingdom. \\
email: saif.ahmad@lis-lab.fr, seifeddine.benelghali@lis-lab.fr, hafiz.ahmed@sheffield.ac.uk}
}

\author{\IEEEauthorblockN{Saif Ahmad}
\and
\IEEEauthorblockN{Seifeddine BEN ELGHALI}
\and
\IEEEauthorblockN{Hafiz Ahmed}
}
\IEEEoverridecommandlockouts
\maketitle
\IEEEpubidadjcol
\begin{abstract}

In the context of dynamic virtual power plants (DVPPs), the integration of frequency containment reserve (FCR) and fast frequency control (FFC) enabled via local compensation of power imbalance represents a significant advancement in decentralized frequency regulation. However, they still have to cope with the limited power and energy capacities associated with commonly available storage solutions. This work combines a disturbance estimation based decentralized local control with distributed imbalance compensation in the event of local shortfall. The layered architecture facilitates fast local corrections in power setpoints while enabling coordination between neighbouring DVPP nodes to leverage the aggregated capacity, ensuring scalable and efficient operation suitable for renewable-heavy future grids. The proposed approach is validated on an illustrative 4-bus system with a high percentage of renewables.

\end{abstract}


\section{Introduction}
Power systems are currently undergoing rapid transformation to address the limitations and adverse environmental impacts of conventional fossil fuel-based energy sources with distributed generators comprising \textit{renewable energy sources} (RESs) such as solar and wind providing a promising solution to this problem. However, uncoordinated integration of renewables in the electricity grid results in issues such as large voltage/frequency swings, increased operational costs (\textit{e.g.} payouts to balancing market entities, spinning reserve costs etc.) to compensate for inaccurate generation/load forecasts and more frequent blackouts due to generator and/or transmission network overloads \cite{blackout2018australia}. To ensure increased penetration of renewables in the electricity grid and to meet the increasing energy demand with intermittent and unpredictable RESs, controlled aggregation of \textit{distributed RESs }(DRESs) in the form of \textit{microgrids} (MGs) \cite{justo2013ac} and \textit{ virtual power plants} (VPPs) \cite{naval2021virtual}, energy communities etc. have emerged as a powerful tool to balance the uncertainty in production and allow small to medium capacity generators as well as prosumers to monetize their flexibility by participating in the energy, capacity and flexibility markets thereby generating community interest \cite{SeifVPPMarketsModelsOptChallengesOpportunities}. Furthermore, it is expected that a bulk of the ancillary services in the future power grids will have to be shouldered by non-synchronous DERs, implying the significance of these aggregated systems in ensuring grid stability and reliability. 

In particular, the concept of \textit{dynamic virtual power plants} (DVPP) has been proposed to pave the way for DERs to provision future ancillary services \cite{dvpp2022florian}.  In this work, we focus on leveraging the potential of DVPPs for frequency regulation in renewable-heavy power grids. The absence of rotational kinetic energy in DRESs and heterogeneous inertia distribution in the network results in faster frequency dynamics, thereby making the task all the more challenging \cite{lowInertiaChallenges2018}. On the other hand, when controlled properly, VPPs can be leveraged to provide inertia support via grid-forming IBRs to reinforce the system response in the event of contingencies.

Traditionally, frequency regulation was implemented in a centralised manner via \textit{automatic generation control} (AGC) where a central authority generates regulatory signals. Despite being effective for decades, this approach is inherently limited in response speed, making it unsuitable for fast frequency regulation \cite{dataDrivenFastFreqControlJWSporco}. 
In recent years, several studies have investigated the impact of increased penetration of RESs into the grid infrastructure and proposed IBR-based frequency regulation schemes, some of which are briefly discussed here. Model-free learning based techniques, such as those based on reinforcement learning (RL) \cite{rlFreqContrl2018,rlFreqCtrl2022}, are usually without any stability certificates and work by imposing soft penalties for deviations outside a predefined range, which restricts their application on real-time systems. Furthermore, RL-based techniques require massive training datasets, which compounds implementation complexity. On the other hand, model-based techniques such as model predictive control can satisfy hard constraints during transients, however, this capability relies heavily on the accuracy of the system model and places significant communication overheads \cite{mpc,HiererchicalfastfreqControlJWSporco}. Another model-based approach leveraging disturbance estimator (DE) was introduced in  \cite{HiererchicalfastfreqControlJWSporco} for fast frequency regulation, in combination with distributed optimization based on \textit{alternating direction method of multipliers} (ADMM) to overcome the shortfall in local RESs by coordinating with neighbouring assets.  However, the coordination of neighbouring local area controllers was implemented through a central coordinator which significantly increases the communication overhead and impacts scalability for large systems.

In this work, we follow the general framework adopted in \cite{HiererchicalfastfreqControlJWSporco}, where the power system is partitioned into smaller areas (DVPP nodes in our case) having unmeasured active power
and then a controller is designed to coordinate frequency regulation in the aggregated system. In contrast to  \cite{HiererchicalfastfreqControlJWSporco, dataDrivenFastFreqControlJWSporco}, the disturbance estimator in our work allows a more general internal model for the unmeasured active power while the coordinating layer is completely distributed, requiring communication only between neighbouring nodes. This, in turn, improves the system's reliability and preserves data privacy.

The remaining sections in this paper are organised as follows: Section \ref{section:model} introduces the modeling of electrical network and different nodes using the swing equation, followed by the design of a disturbance estimation mechanism for local control and distributed coordination in Section \ref{section:proposed approach}. Section \ref{section: simulation} presents a numerical study to evaluate the efficacy of the proposed approach and conclusions are drawn in Section \ref{section:conclusion}.

\subsection{Notations:} $\bm{I_n}$ is an identity matrix of dimension $n\times n$, $\textbf{0}$ is a zero vector or matrix of appropriate dimension, $\textrm{diag}(a_1,\dots,a_n)$ denotes a diagonal matrix with $a_1,\dots,a_n$ as diagonal elements. Given a symmetric matric $\textbf{M}$, $\textbf{M} \prec \textbf{0}$  means that $\textbf{M}$ is negative-definite. The notation for positive definiteness is analogous.


\section{Mathematical Model}\label{section:model}
We consider the electrical grid to be composed of a network of interconnected nodes, where the local frequency dynamics of each node is defined by a swing equation. Nodes can be of different types depending on the constituting elements and available regulation mechanisms. In this work, we distinguish between two types of nodes, which are discussed in the following subsections. The terms `nodes' and `buses' are used interchangeably throughout the paper, with nodes being used mostly in the modeling and algorithmic part and buses in the numerical validation part to link with the physical system.

\subsection{Dynamic Virtual Power Plant Nodes}
 The local frequency dynamics of the $i^{\text{th}}$ DVPP node can be defined by a standard swing equation of the form \cite{swingEqnNature}
\begin{equation}
    \begin{split}
        \dot{\theta}_i&=\wi\\
        \derivativewi&= \dfrac{1}{2\inertiaconstanti}\powerimbalancei
    \end{split}
    \label{eq_swing_model}
\end{equation}
where $\wi$ is the frequency deviation (p.u.), $\inertiaconstanti$ is inertia constant (seconds) and $\powerimbalancei$ is the net power imbalance (p.u.). The DVPP nodes are considered to be without synchronous generators (SGs), and are characterized by low inertia and damping provided mostly by IBRs and rotating machines such as large motors. The power imbalance can be expressed as
\begin{equation}
    \begin{split}
            \powerimbalancei=-\pielec+\unmeasuredpoweri+\pirenewable-\actualpifcr -\actualcontrolsignal-\sum_{j}\pflow{i}{j}
    \end{split}
    \label{eq_power_imbalance}
\end{equation}
where $\pielec$ is electrical load, $\unmeasuredpoweri$ is the unmeasured active power injected, $\pirenewable$ is the power injected by RESs including storage, $\actualpifcr$ is power injected by the assets participating in the frequency containment reserve (FCR) (discussed in Section \ref{section:fcr}), $\actualcontrolsignal$ is the power consumed by the \textit{battery energy storage system} (BESS) and $\sum_{j}\pflow{i}{j}$ is the total power flowing out of the $i^{\text{th}}$ node via tie-lines which depends on the voltage phase angle ($\theta_i,\theta_j$) at the nodes (or rather the difference between the two) and is given by
\begin{equation}
    \pflow{i}{j}=\dfrac{1}{\reactancei}\sin{(\theta_i-\theta_j)},
    \label{eq_power_flow}
\end{equation}
where $\reactancei$ is the reactance between nodes $i$ and $j$.
We consider that the power flowing into and out of the DVPP via tie lines is measured, along with the power generated/consumed by all the assets/loads except for a portion of the load ($\unmeasuredpoweri$) which is not measured by the smart meters. It is also assumed that the net power imbalance is zero initially $(t=0 \ s)$ i.e. generation $(\pirenewable)$ is equal to consumption $(\pielec)$, which cancels their net effect on the frequency dynamics. Hence, all frequency deviations, power values and control signals are zero initially, which simplifies the simulation scenario.

The BESS dynamics is defined as 
\begin{equation}
     \derivativeactualcontrolsignal=\begin{cases}
         \dfrac{1}{\ibrtimeconstant}(\idealcontrolsignal-\actualcontrolsignal) &; \ \ \actualcontrolsignal\in\left[\piibrmin, \piibrmax\right] \\
         0 &;  \ \ \text{otherwise}
     \end{cases}
    \label{eq_RES_dynamics}
\end{equation}
where $\ibrtimeconstant$ is the time constant imposed by the local device level control-loop, $\idealcontrolsignal$ is the setpoint for BESS,  $\piibrmin$ and $\piibrmax$ are device power limits imposed either due to physical limitations, BMS behaviour, or to have a power buffer to comply with accepted capacity bids.
\subsubsection{FCR Response}\label{section:fcr}
Frequency containment reserve (FCR) are used to contain the frequency deviations under normal operation between 49.9 to 50.1 Hz (FCR-N) and in the event of large disturbances up to 49.5 and 50.5 Hz (FCR-D).  In this work, we only consider FCR-N where cumulative active power response in p.u. of the contracted FCR-N assets can be expressed as 
\begin{equation}
    \pifcrn = \begin{cases}
\begin{split}     
        \gainfcrn \wi   ;& \  |\wi|\leq\rangefcrn\\
        \pifcrncapacity\wi \quad  ;& \ |\wi|>\rangefcrn
\end{split}
    \end{cases}
\end{equation}
where the range of FCR-N is denoted by $\rangefcrn=0.2\pi/50$ in p.u., $\gainfcrn=\frac{\pifcrncapacity}{\rangefcrn}$ and $\pifcrncapacity$ is the contracted FCR-N capacity. Once an FCR bid is accepted, that capacity needs to be reserved (for activation by TSO when needed), and cannot be traded again in concurrent or subsequent day-ahead markets. It is assumed that the assets participating in FCR have metered connections to facilitate reconciliation. 



\subsubsection{Radom Variations in $\pielec$ and $\pirenewable$}\label{section: random variations in load and res}
 Random variations in the electrical load due to On/Off switching of medium to low power appliances can be modeled as a sum of scaled (in magnitude) \textit{pseudo random binary sequences} (PRBSs) 
 \begin{equation}
 \begin{split}
     \nu_n&=\sum_{k=1}^{N_{\nu}}\nu^{k}_n,\\
\nu^{k}_0 &= a^k\\
\nu^k_n &=
\begin{cases}
-\nu^{k}_{n-1}, & \text{with probability } \mathcal{P}_{k,n}>p_k,\\
\nu^{k}_{n-1}, & \text{otherwise},
\end{cases}
\end{split}
\label{eq:prbs}
 \end{equation}
 where $n$ denotes the discrete time instant, $N_{\nu}$ is the number of PRBSs used to generate the load variations, $\mathcal{P}_{k,n}\sim\mathrm{Uniform}(0,1) \text{ sampled independently at each step}, \ p^k=k/s_f \text{ is the flip probability per step}, \ a^k=h/d^k, \  d^k=\begin{cases}
         1   ,& k=1\\
         2(k-1)   ,& k\geq2
 \end{cases} $, $h$ scales the magnitude of PRBS and $s_f$ controls the probability of how frequent the switching occurs. 

 On the other hand, random variations in the power generated by the renewables is modeled via  \textit{Brownian motion with soft reset} (BMR) 
 given by 
 \begin{equation}
     \begin{split}
        \nu_n^{\star}&= \nu_{n-1}^{\star}+\sigma\sqrt{\Delta t}\xi_n\\
         \nu_n^{\text{RES}}&=\begin{cases}
             (1-\lambda_r)\nu_n^{\star}, & |\nu_n^{\star}| > \theta_{\text{TH}},\\
\nu_n^{\star},              & \text{otherwise},
         \end{cases}
     \end{split}
     \label{eq:brownian_with_reset}
 \end{equation}
 where $\theta_{\text{TH}}$ is the reset threshold, $\xi_n\sim\mathcal{N}(0,1)$ is sampled independently at each time-step, $\Delta t$ is the step size, $\sigma$ is the diffusion intensity and $\lambda_r$ is the decay factor. The reset helps in avoiding runaways and keeps the value within limits.
 
\subsection{Synchronous Generator Nodes}
We assume that synchronous generators (SGs) are only capable of providing  droop support and are characterised by the following second-order dynamics:
\begin{equation}
    \begin{split}
        \derivativewi&= \dfrac{1}{2\inertiaconstanti}\left(\powerimbalancei-\dfrac{\wi}{\ibrprimarygain}\right)\\
        \powerimbalancei&= \pimech-\pielec-\sum_{j}\pflow{i}{j}\\
        \derivativepimech&=\dfrac{1}{\governortimeconstant}\bigg(-\dfrac{\wi}{\droopcoefficientsg}-\pimech\bigg)
    \end{split}
    \label{eq:model_synchronous_generator}
\end{equation}
 where $\pimech$ is the mechanical power input,  $\droopcoefficientsg$ and $\ibrprimarygain$ are the primary control gains of IBRs and SG, respectively, $\governortimeconstant$ is the turbine-governor time-constant, while $\ibrprimarygain,\powerimbalancei,\pielec$ and $\pflow{i}{j}$ are the same as in \eqref{eq_swing_model} and \eqref{eq_power_imbalance}. In addition to the mechanical power, a major difference between DVPP nodes and SG nodes is that the inertia constant for the latter is typically much higher compared to DVPP which only contains renewables.



\section{Proposed Approach}\label{section:proposed approach}
The main idea behind our approach is to first locally compensate any power imbalance at the node by available RESs + storage and second, to coordinate neighbouring DVPP nodes to compensate the imbalance in the event of local shortfall. 
To accomplish this task, it is essential to estimate the unmeasured active power, which can cause local frequency deviation. In this section, we estimate the unmeasured active power by introducing state augmentation based on an assumed internal model for $\unmeasuredpoweri$. 
\begin{assumption}
    The unmeasured active power is generated by a linear exo-system
    \begin{equation}
        \begin{split}
            \bm{\dot{\zeta}_i}&=\bm{A\zeta_i}\\
            \unmeasuredpoweri&=\bm{C\zeta_i},
        \end{split}
        \label{eq:dist_internal_model}
    \end{equation}
    \label{assumption1}
\end{assumption}
where $\bm{A}\in\mathbb{R}^{m\times m}, \ \bm{C}\in\mathbb{R}^{1\times m}$ and $\bm{\zeta_i}=\mathbb{R}^{m\times 1}$ is the state vector for exo-system.

The above assumption is a generalisation of the commonly used $\derivativeunmeasuredpoweri=0$  \cite{dataDrivenFastFreqControlJWSporco,HiererchicalfastfreqControlJWSporco} and allows the flexibility to consider a wide range of linear dynamics such as sinusoidal or time polynomial function for  $\unmeasuredpoweri$ \cite{ahmad2021active,ahmad2023TIA}. Following \textit{Assumption} \ref{assumption1}, the  state-space model for the augmented system  can be defined as
\begin{equation}
        \bm{\dot{x}}=\bm{\mathcal{A}x}+\bm{B}(-\actualcontrolsignal+f),
    \label{eq:augmented_model}
\end{equation}
where $\bm{x}=[\wi, \ \bm{\zeta_i^T}]^T$ is the augmented state vector, $f=-\pielec+\unmeasuredpoweri+\pirenewable-\actualpifcr -\sum_{j}\pflow{i}{j}$ denotes the measured quantities, whereas state and input matrices are respectively given by,
$$\bm{\mathcal{A}}=\begin{bmatrix}
    0 & \theta_i\bm{C}\\
    \bm{0} & \bm{A}
\end{bmatrix}, \ \bm{B}=\begin{bmatrix}
    \theta_i\\ \bm{0}
\end{bmatrix}, \theta_i=\dfrac{1}{2\inertiaconstanti}.$$

Consequently, a Luenberger-type disturbance estimator (DE) can be designed for the augmented model in \eqref{eq:augmented_model}:
\begin{equation}
\begin{split}
    \bm{\dot{\hat{x}}}&=\bm{\mathcal{A}\hat{x}}+\bm{B}(-\actualcontrolsignal+f)+\bm{\kappa}(\wi-\hatwi)\\
    \hatwi&=\bm{\mathcal{C}\hat{x}}
\end{split}
    \label{eq:estimator}
\end{equation}
where $\bm{\mathcal{C}}=\begin{bmatrix}
    1, \bm{0}
\end{bmatrix}$ is the output matrix, $\kappa\in\mathbb{R}^{m+1}$ is the estimator gain vector and variables with a circumflex denote the estimate of original quantities. It is to be noted that the DE in \eqref{eq:estimator} also estimates the local frequency $(\hatwi)$ in addition to the unmeasured active power, which can be used for power modulation in case of FCR to suppress the effect of measurement noise and ensure grid compliance.  Unlike \cite{dataDrivenFastFreqControlJWSporco,HiererchicalfastfreqControlJWSporco}, we are not concerned with local mismatched power sharing among IBRs and only focus on coordination among DVPP nodes to balance demand with generation, which will be discussed later in Section \ref{section: distributed control}.

   Clearly, the designed DE in \eqref{eq:estimator} relies on model information, particularly the inertia constant $(\inertiaconstanti)$ of the DVPP node.  However, the designed estimator is robust towards model uncertainty in $\inertiaconstanti$ to some extent and returns an accurate estimate of $\unmeasuredpoweri$ despite model mismatch, if the internal model for the exo-system is accurate enough and the following LMI is satisfied
   \begin{equation}
       \bm{P}(\bm{\mathcal{A}}-\bm{\kappa\mathcal{C}})+(\bm{\mathcal{A}}-\bm{\kappa\mathcal{C}})^T\bm{P}\prec 0
       \label{eq:lmi_stability}
   \end{equation}
where $\bm{P}=\bm{P^T}\succ 0$. The above LMI can be obtained following standard Lyapunov argument (refer \cite{ahmad2024TIA} for more details).

For the sake of simplicity and keeping in line with recent literature, we consider the simple internal model of $\derivativeunmeasuredpoweri=0$ in this work, which results in 
$\bm{A}=\bm{0}$ and $\bm{C}=1$.
\subsection{Decentralised Local Regulation}\label{section: local control}
The setpoint for BESS i.e. $\actualcontrolsignal$ is selected as
\begin{equation}
\begin{split}
        \idealcontrolsignal=-\pielec+\hatunmeasuredpoweri+\pirenewable,
\end{split}
\label{eq:local_control_signal_ideal_dyna}
\end{equation}
to compensate for the net internal power mismatch within a DVPP node in real-time, based on the estimate $\hatunmeasuredpoweri$ obtained from DE in \eqref{eq:estimator}. The idea here is to cancel out all the internal power imbalances (except for the FCR which introduces much needed damping) so that the RoCoF and $\sum_j\pflow{i}{j}$ become zero for the DVPP node.  For a system composed entirely of DVPP nodes, the control action in \eqref{eq:local_control_signal_ideal_dyna} can stabilise the entire system if $\idealcontrolsignal\in[\piibrmin,\piibrmax]$ despite being decentralised. However, when $\idealcontrolsignal\notin[\piibrmin,\piibrmax]$, the local resources are insufficient to compensate for the power imbalance and require assistance from neighbouring nodes, which is discussed next. 

\subsection{Distributed Redispatch to Overcome Local Shortfall}\label{section: distributed control}
We briefly introduce the concept of a graph and associated theory which will be used to define the communication network between nodes. A graph \( \mathcal{G} = (\mathcal{V}, \mathcal{E}) \) is defined as a set of vertices (or nodes) \( \mathcal{V} = \{1, 2, \dots, N\} \) and a set of edges \( \mathcal{E} \subseteq \mathcal{V} \times \mathcal{V} \), where each edge \( (i,j) \in \mathcal{E} \) represents a connection between two nodes \(i\) and \(j\).  We consider an undirected graph which implies that the communication between any two nodes is bidirectional.

 As defined earlier, $\idealcontrolsignal$ is the required power to be consumed by the BESS (set-point regulated by the local controller), and $\actualcontrolsignal$ is the measured actual power consumption by the BESS. Assuming $\actualcontrolsignal\neq\idealcontrolsignal$ due to physical or synthetic limitations imposed due to market participation, the mismatch can be compensated via a distributed averaging proportional integral (DAPI) controller \cite{dapiInceptionJWSPorco2013Automatica,dapiStabilityJWSPorco2020IeeeCSS} of the form
\begin{equation}
\derivativedistributedcontrolsignali=-\distributedcontrolgaini\left[\powerimbalancedisti+\sum_{j\in\neighbourhoodi}\communicationweightj\left(\dfrac{\distributedcontrolsignali}{\scalingfactordistributed{i}}-\dfrac{\distributedcontrolsignalj}{\scalingfactordistributed{j}}\right)\right]
\label{eq:dapi}
\end{equation}
where $\powerimbalancedisti=-\pielec+\hatunmeasuredpoweri+\pirenewable -\actualcontrolsignal$, $\distributedcontrolgaini \text{ is a positive gain, } \communicationweightj= 1$ if $(i,j)\in\mathcal{E}$ else $0$, while $\scalingfactordistributed{i},\scalingfactordistributed{j}$ are scaling factors that control the contribution of a specific node in the event of local shortfall at any node. A communication delay of $\communicationdelay \ s$ is included in the computation of $\distributedcontrolsignali$ by delaying $\distributedcontrolsignalj$ from all the neighbouring nodes.  The final control law adds the distributed control action to the local control law in \eqref{eq:local_control_signal_ideal_dyna} 
\begin{equation}
\idealcontrolsignal=\distributedcontrolsignali-\pielec+\hatunmeasuredpoweri+\pirenewable.
\label{eq:final_control}
\end{equation}


\section{Numerical Simulation} \label{section: simulation}
For the numerical study, we consider a 4-bus system to test the efficacy of our approach. All simulations are performed in Python (ver. 3.11), where the equations are implemented using the explicit forward Euler method for numerical integration with a step-time of 5e-4 s.
Two scenarios are considered to evaluate different operational aspects. The parameters used in numerical analysis are listed in Table \ref{table:parameters}, and the setting for all the scenarios is defined  as follows:

\begin{table}[!t]
  \centering
  \begin{tabularx}{\columnwidth}{|c|X|}
    \hline
    \textbf{Type} & \textbf{Parameters}\\
    \hline
    DVPP node&
      $\inertiaconstanti\in[0.01, 0.1]\,\mathrm{s},\; \scalingfactordistributed{i}=\{1,2,3\}, \; \distributedcontrolgaini=1, 
       \bm{\kappa}=[20, \ 100]^T, \; \communicationweightj =1 \text{ if $(i,j)\in\mathcal{E}$ else $0$, $\pifcrncapacity=\{0.005,0.003,0.001\}, \;$} \;$ $ \communicationdelay=0.5 \ s, \ibrtimeconstant=0.1 \ s$ \\
    \hline
    SG node&
      $\inertiaconstanti=\{4, \ 0.005\}\,\mathrm{s},\;
       \ibrprimarygain=0.05,\;
       \governortimeconstant=2 \ s$\\
    \hline
    $\{\pirenewable, \pielec\}$&$\mathcal{N}_v=8, h=0.002,s_f=1e04, \lambda_r=0.5, \sigma=0.005, \theta_{\text{TH}}=0.02, \Delta t= 5e-04$\\
    \hline
    Network &$\reactance{1,2}=0.1,\reactance{2,3}=0.1, \reactance{3,1}=0.1, \reactance{3,4}=0.02$, $\mathcal{E}:=\{(1,2),(2,3)\}$ \\
    \hline
  \end{tabularx}
  \caption{System parameters}
  \label{table:parameters}
\end{table}
\begin{figure}[]
     \centering
     \begin{subfigure}[b]{0.49\textwidth}
              \centering
         \includegraphics[width=\textwidth]{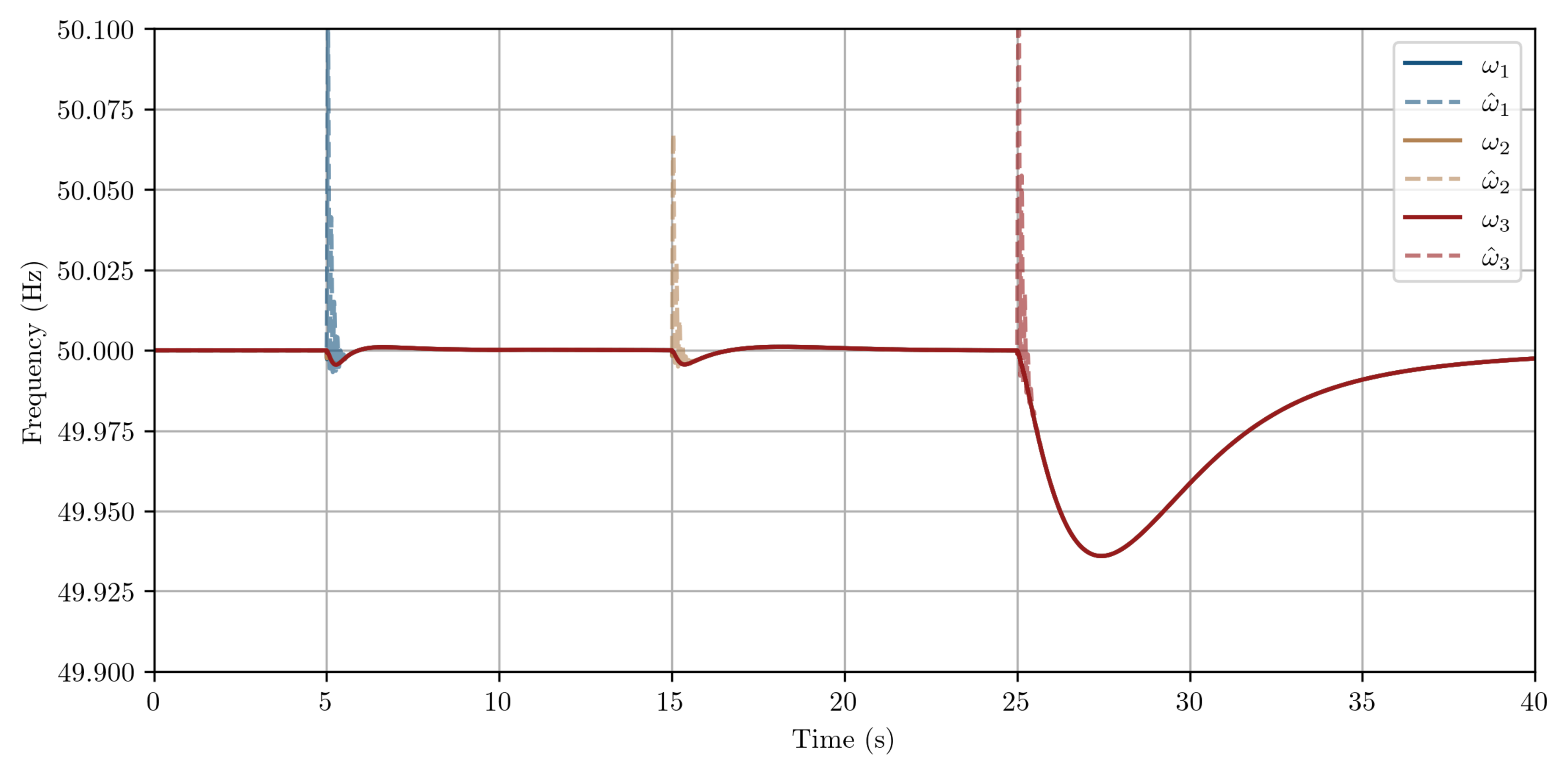}
         \caption{$\wi$ vs $\hatwi$}
         \label{fig:sc1_freq}
     \end{subfigure}
\\
       \begin{subfigure}[b]{0.49\textwidth}

     \centering
         \includegraphics[width=\textwidth]{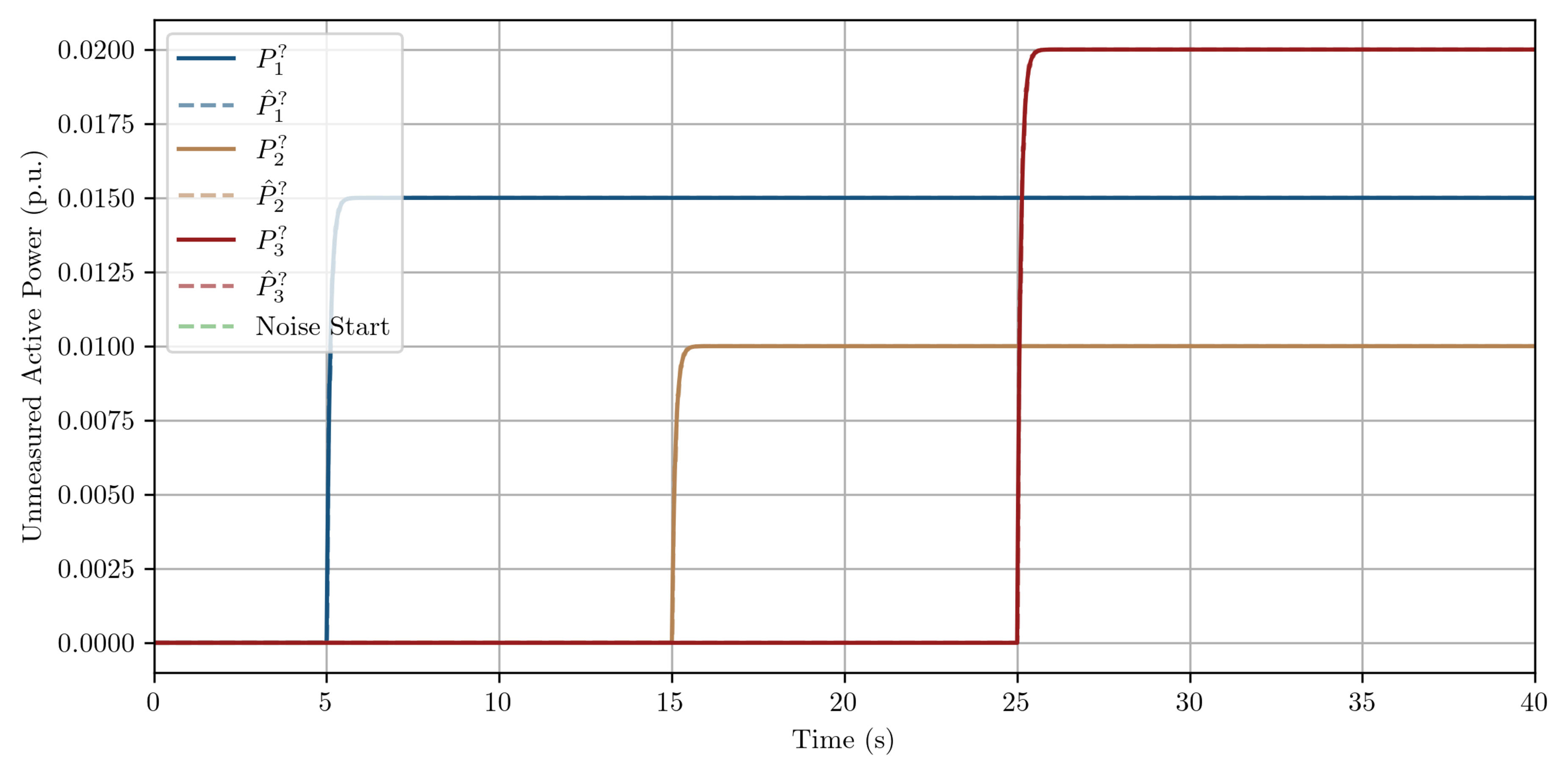}
         \caption{$\unmeasuredpoweri$ vs $\hatunmeasuredpoweri$}
         \label{fig:sc1_unmeasured_active_power}
     \end{subfigure}
     \\
     \vspace{0.2cm}
     \begin{subfigure}[b]{0.49\textwidth}
              \centering
         \includegraphics[width=\textwidth]{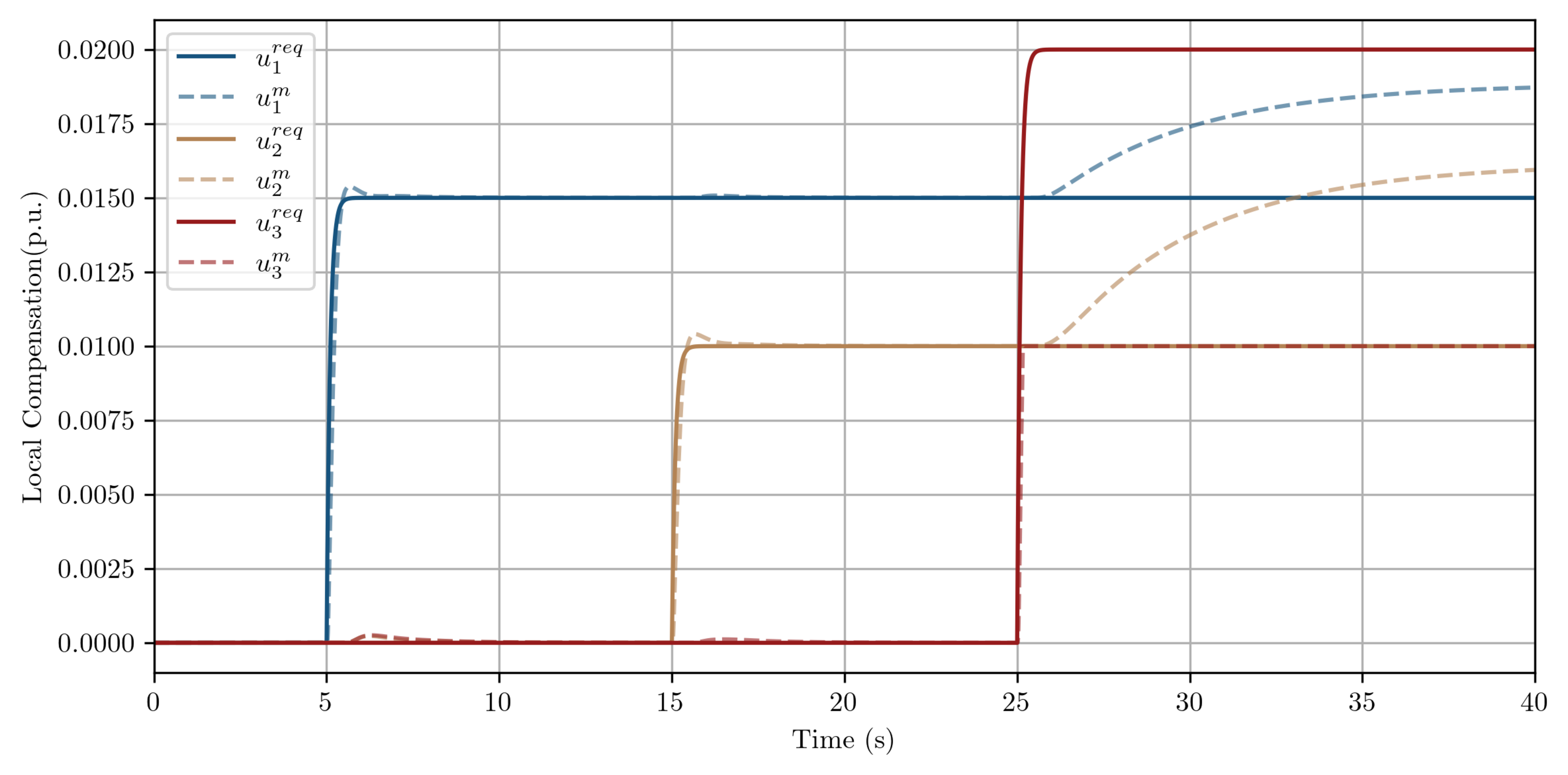}
         \caption{ Required vs measured power supplied by the BESS}
         \label{fig:sc1_net_control}
     \end{subfigure}
\\
       \begin{subfigure}[b]{0.49\textwidth}

     \centering
         \includegraphics[width=\textwidth]{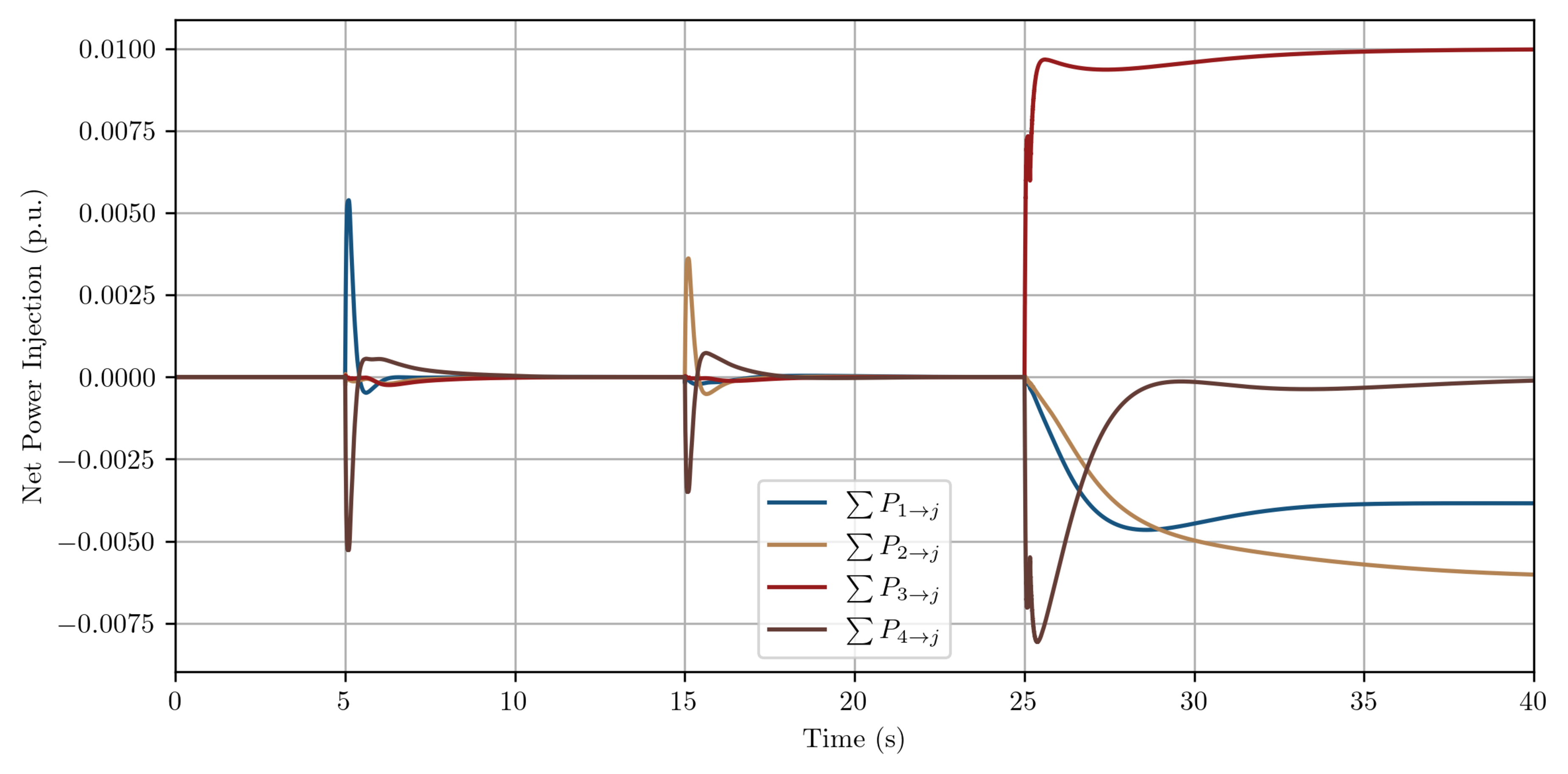}
         \caption{Net tie-line power injection}
         \label{fig:sc1_tieline}
     \end{subfigure}   
        \caption{Scenario 1} 
        \label{fig:scenario1}
\end{figure}

\begin{figure*}[]
     \centering
     \begin{subfigure}[b]{0.49\textwidth}
              \centering
         \includegraphics[width=\textwidth]{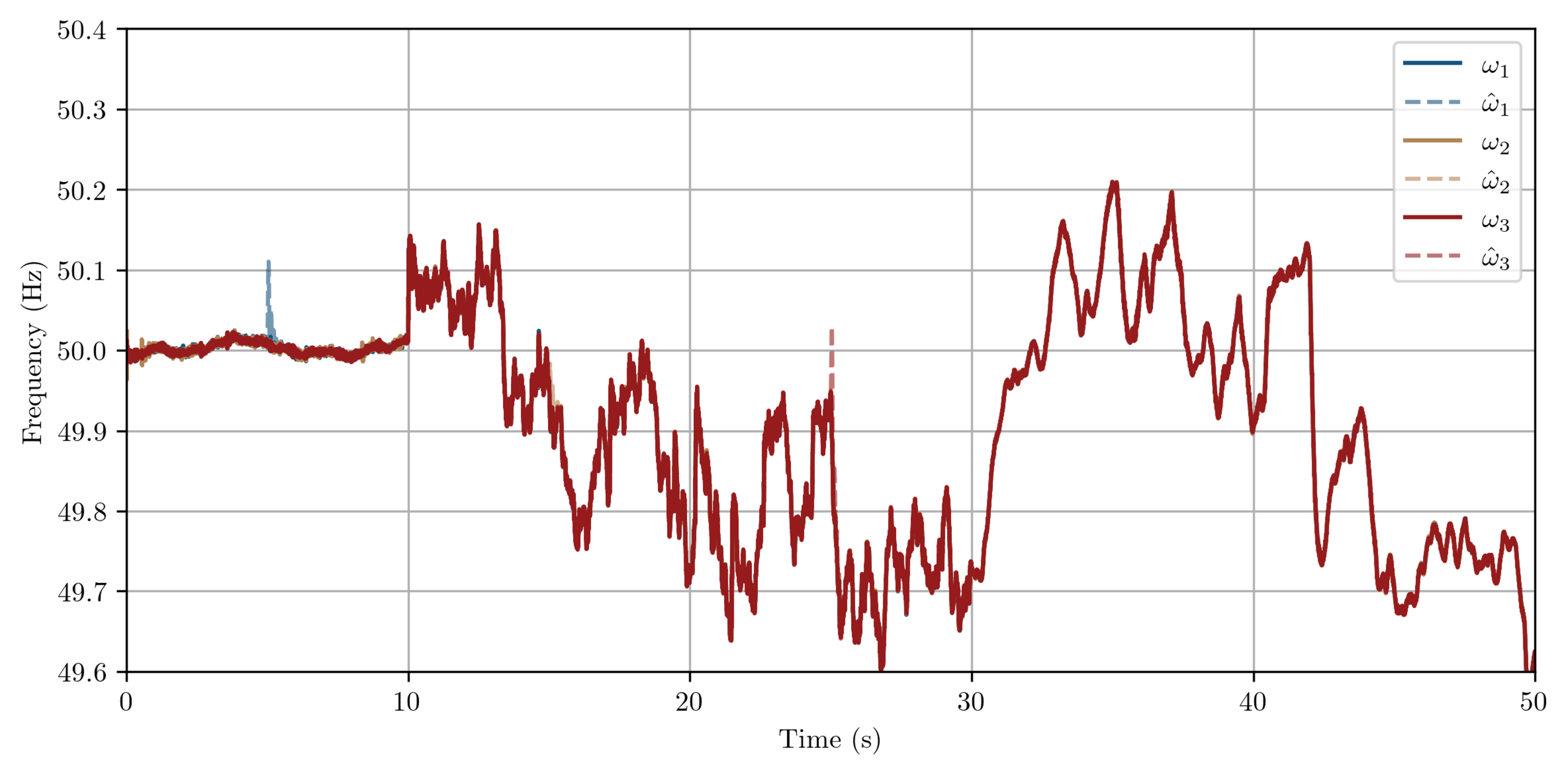}
         \caption{$\wi$ vs $\hatwi$}
         \label{fig:sc1_freq}
     \end{subfigure}
\hfill
       \begin{subfigure}[b]{0.49\textwidth}

     \centering
         \includegraphics[width=\textwidth]{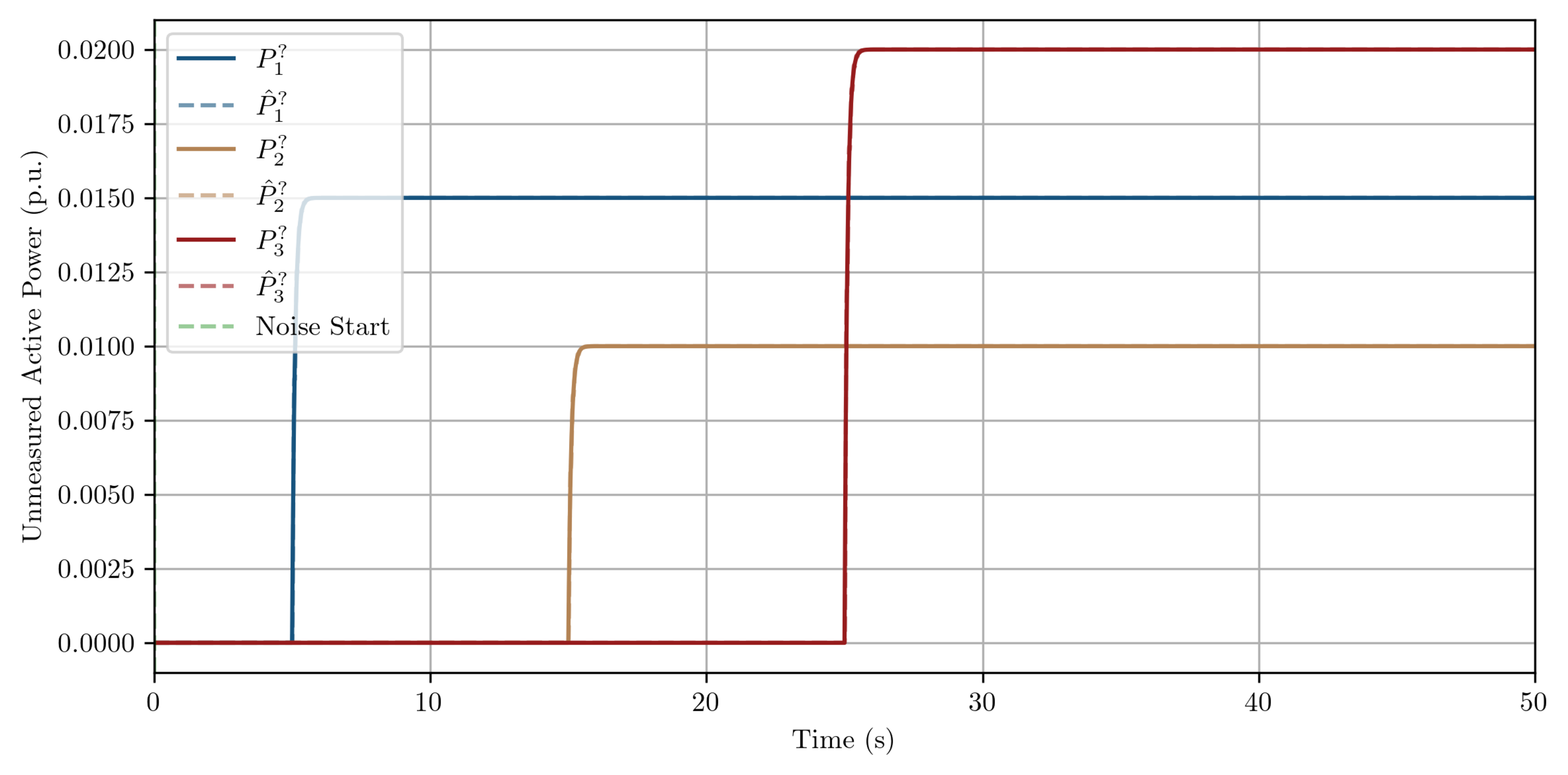}
         \caption{$\unmeasuredpoweri$ vs $\hatunmeasuredpoweri$}
         \label{fig:sc1_unmeasured_active_power}
     \end{subfigure}
     \\
     \begin{subfigure}[b]{0.49\textwidth}

     \centering
         \includegraphics[width=\textwidth]{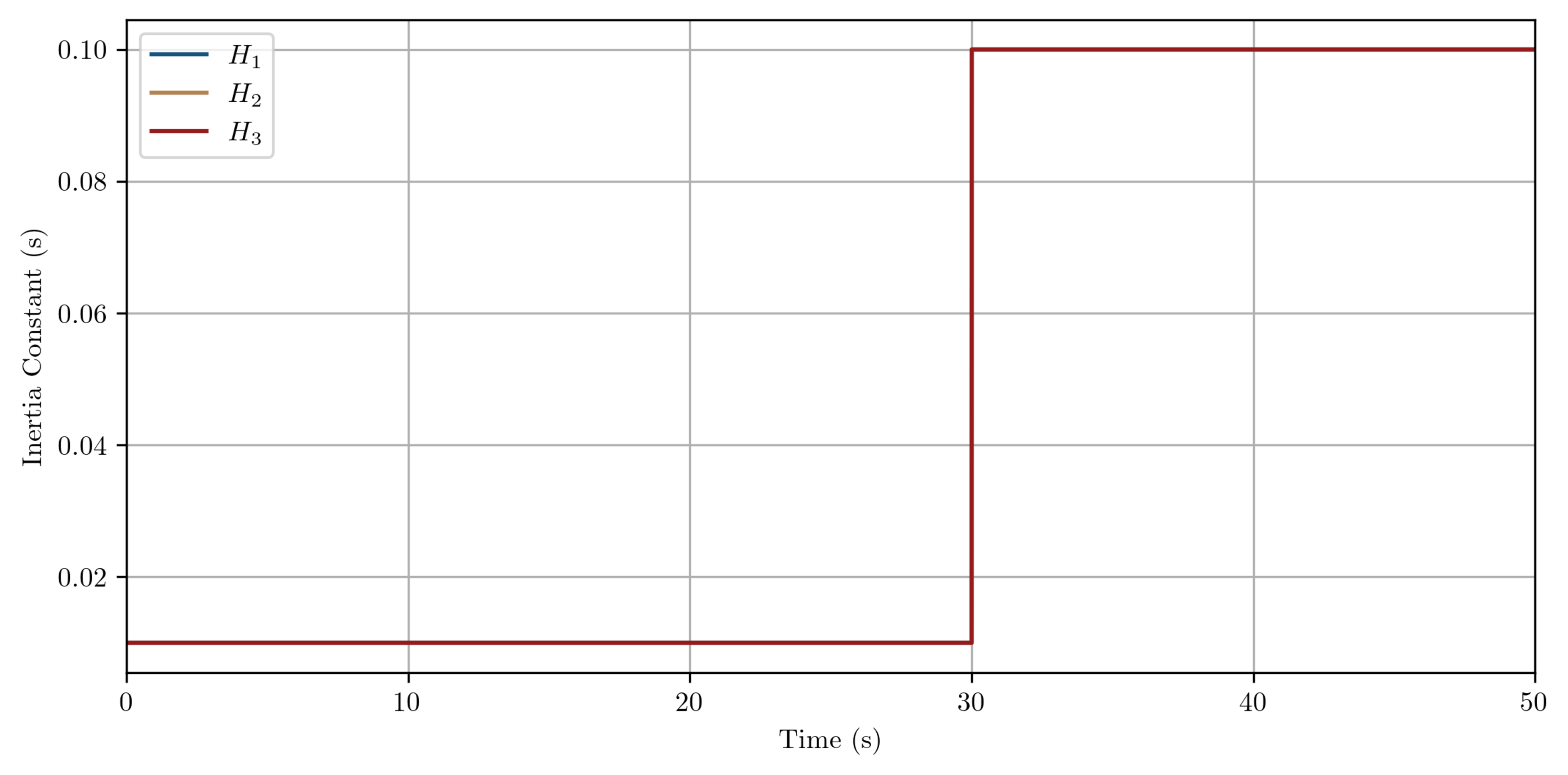}
         \caption{$\inertiaconstanti$}
         \label{fig:sc2_inertia}
     \end{subfigure}
     \hfill
     \begin{subfigure}[b]{0.49\textwidth}

     \centering
         \includegraphics[width=\textwidth]{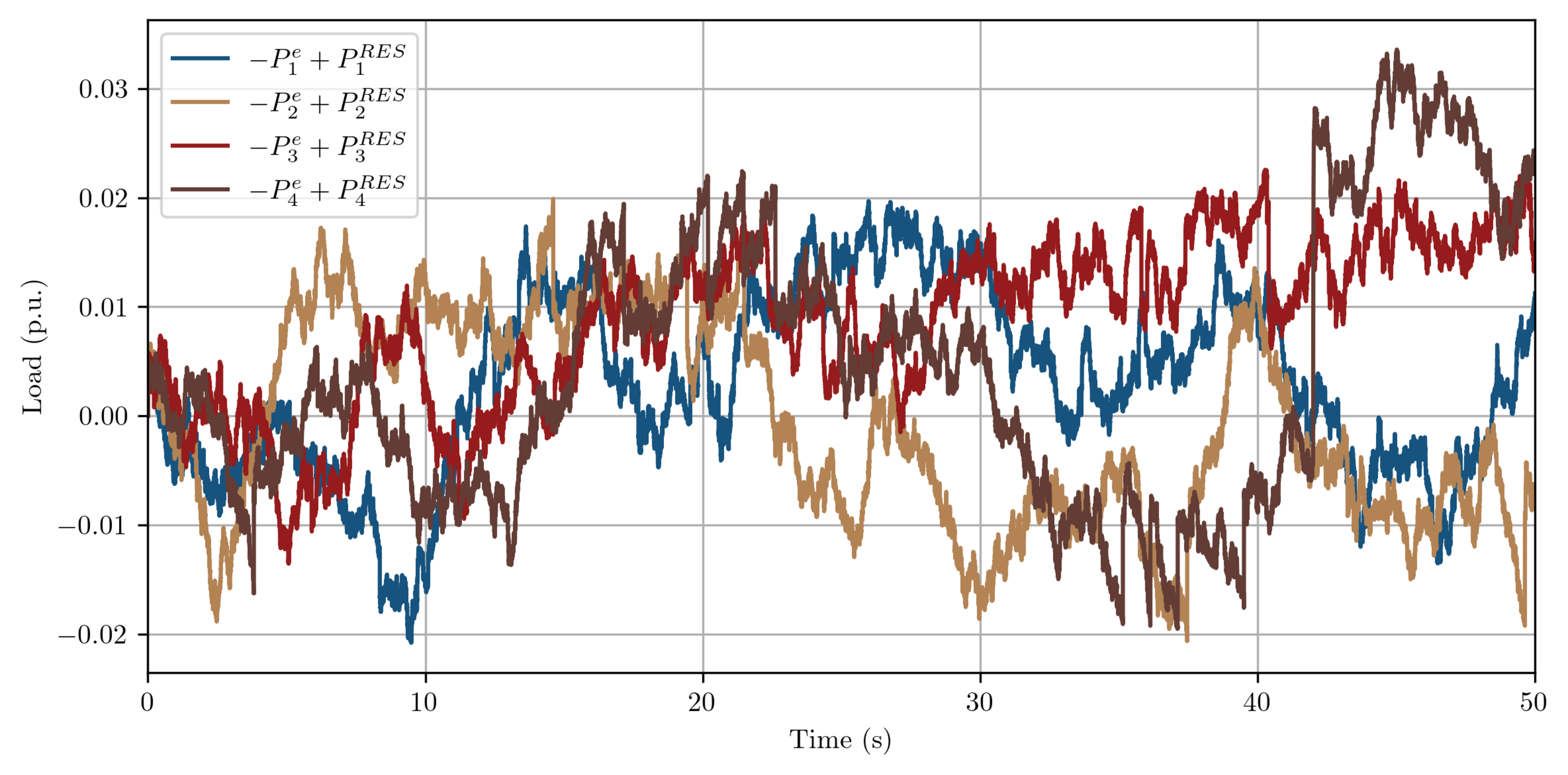}
         \caption{Net effect: $-\pielec+\pirenewable$}
         \label{fig:sc2_load_res}
     \end{subfigure}
     \vspace{0.2cm}
     \begin{subfigure}[b]{0.49\textwidth}
              \centering
         \includegraphics[width=\textwidth]{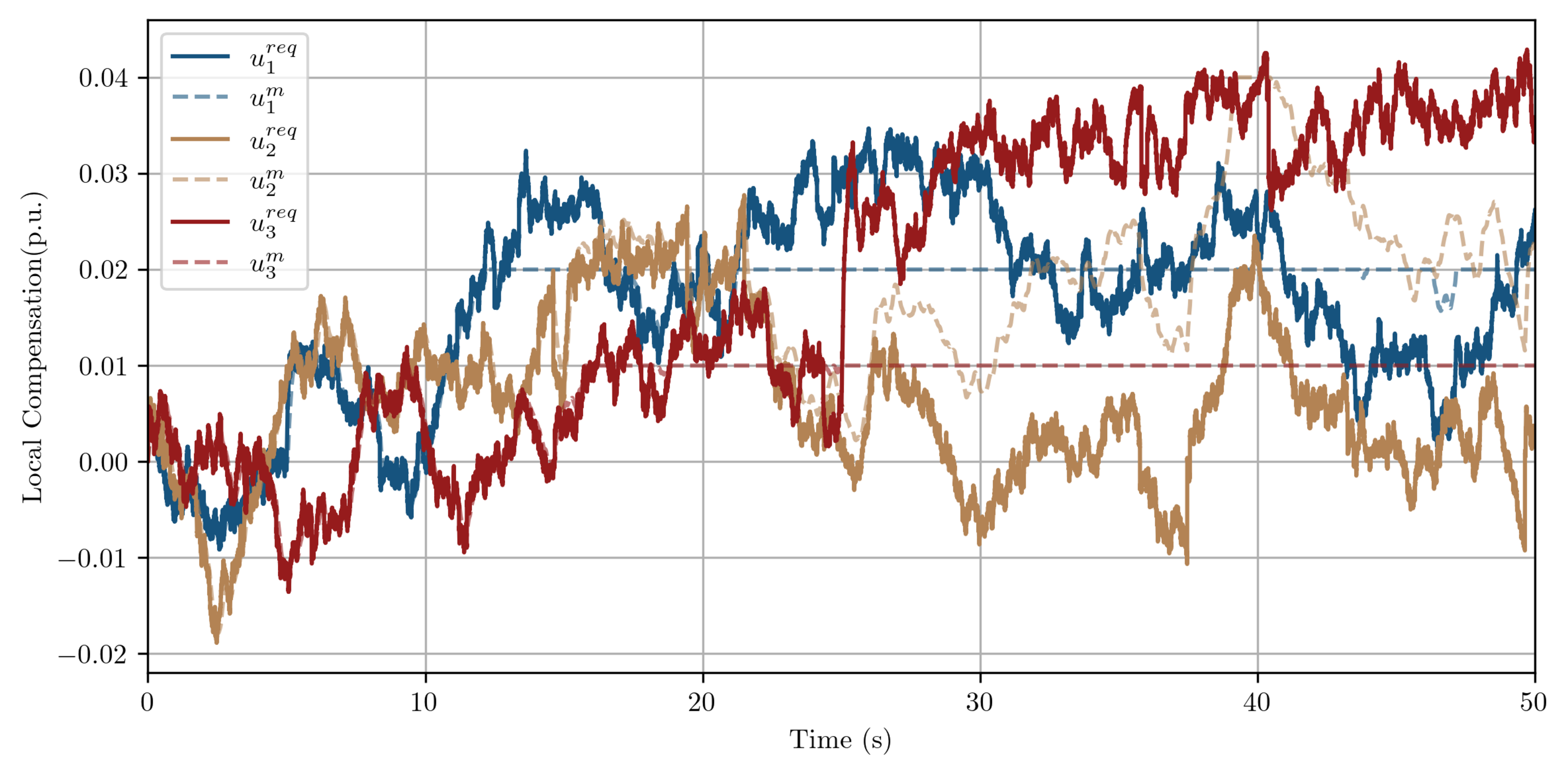}
         \caption{ Required vs measured power supplied by the BESS}
         \label{fig:sc2_net_control}
     \end{subfigure}
\hfill
       \begin{subfigure}[b]{0.49\textwidth}

     \centering
         \includegraphics[width=\textwidth]{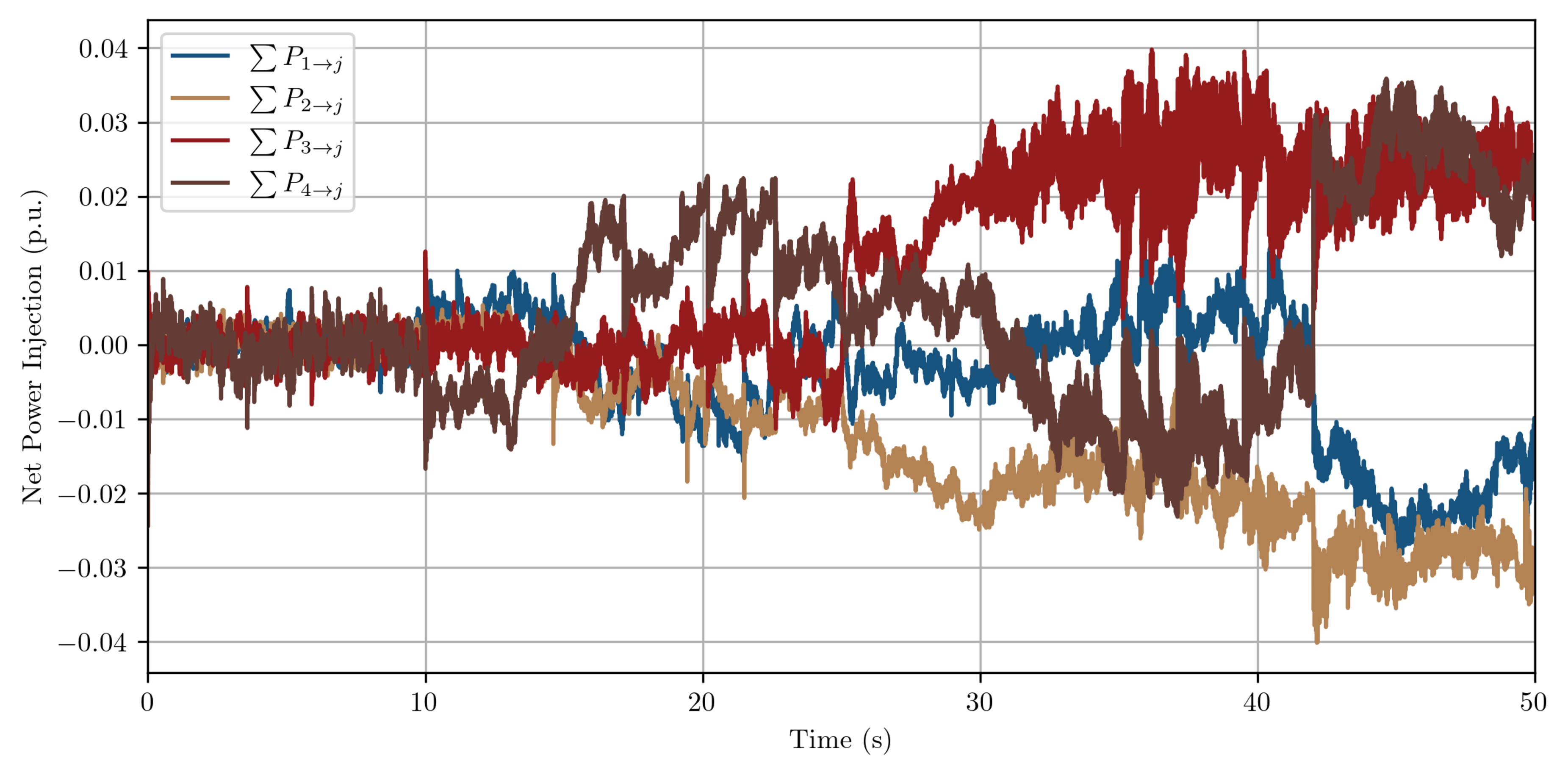}
         \caption{Net power injection via tie-lines $\sum_{j}\pflow{i}{j}$}
         \label{fig:sc2_dist_control}
     \end{subfigure} 
     \\
     \begin{subfigure}[b]{0.49\textwidth}
              \centering
         \includegraphics[width=\textwidth]{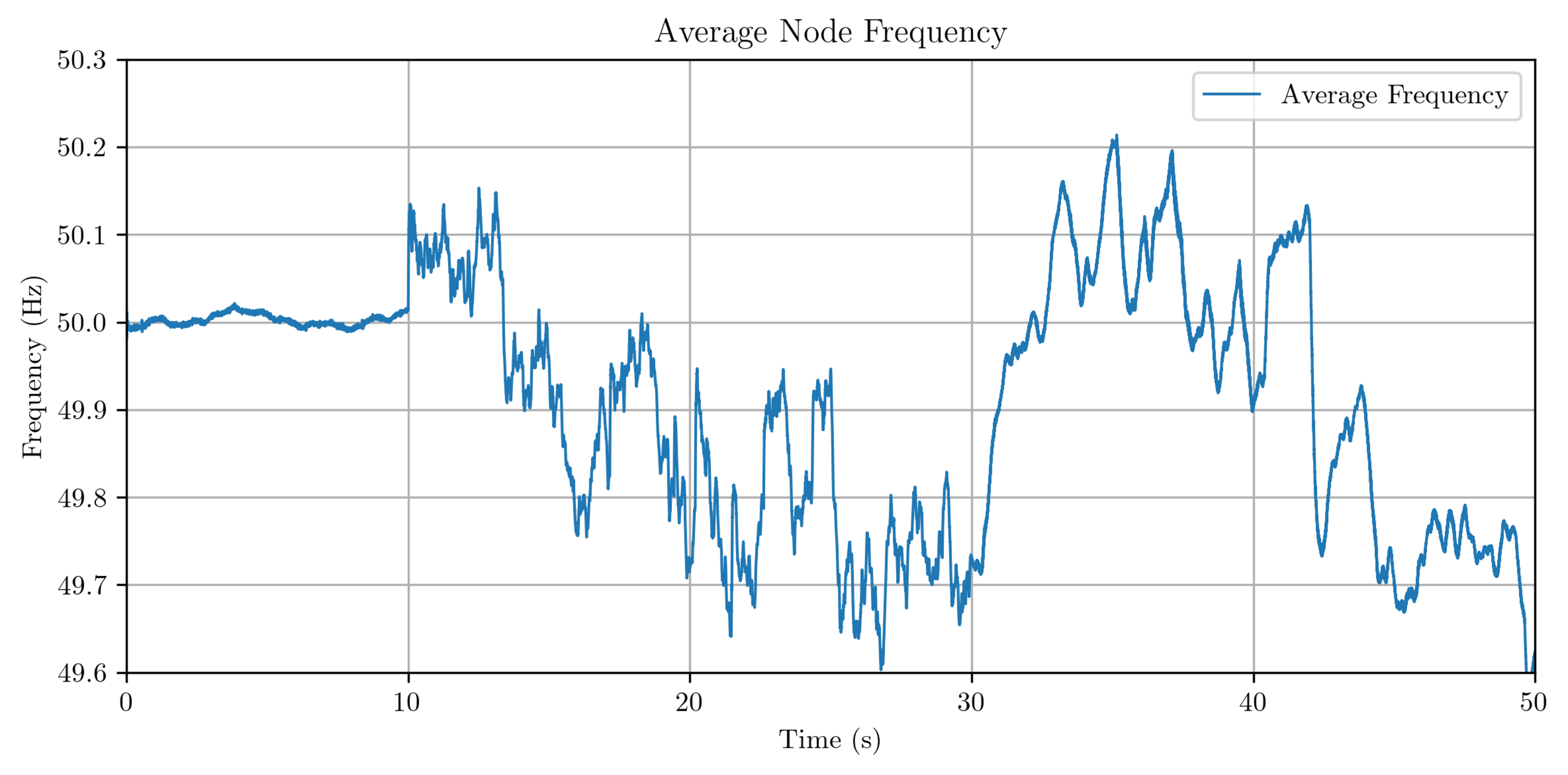}
         \caption{ Average grid frequency with inertia support by the DVPPs.}
         \label{fig:sc2_avg_freq_with_spport}
     \end{subfigure}
\hfill
       \begin{subfigure}[b]{0.49\textwidth}

     \centering
         \includegraphics[width=\textwidth]{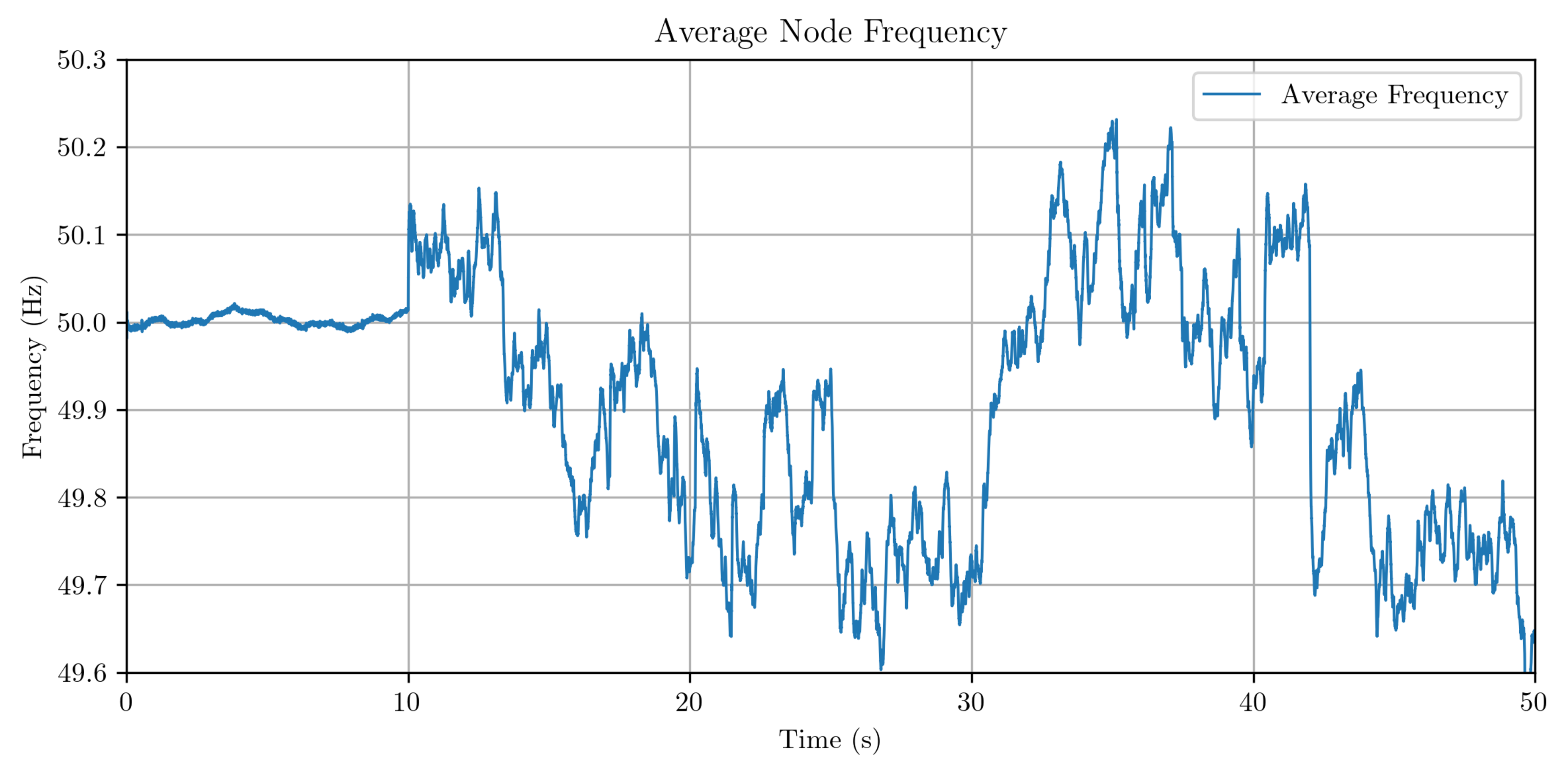}
         \caption{Average grid frequency without inertia support from the DVPPs.}
         \label{fig:sc2_avg_freq_no_spport}
     \end{subfigure}      
        \caption{Scenario 2} 
        \label{fig:scenario2}
\end{figure*}
\begin{itemize}
    \item[\textbf{S1}:] Nodes 1,2, and 3 are DVPPs while node 4 is an SG node. Local resources (generation via RESs and BESS) at some DVPP nodes are insufficient to compensate for the unmeasured active power due to artificial limits set on power converters.  It is assumed that the SG only provides droop support. There are no random variations in RESs and electrical load, while $\unmeasuredpoweri$ undergoes step changes with first-order dynamics having a time constant of $\timeconstantunmeasuredpower$. 
    \item[\textbf{S2}:] Network configuration remains the same as in \textbf{S1}. Local resources are insufficient on some of the DVPP nodes and there are random fluctuations in RESs as well as electrical load. The SG trips during operation, which significantly reduces the inertia and damping of that node.  The DVPP nodes simultaneously increase their inertia following the generator tripping. The performance is compared to a case where the DVPP inertia remains constant. 
\end{itemize}
\subsection{Results and Discussions} The 4-bus network can be interpreted as an island grid connected to a synchronous area via high-capacity transmission lines. The island, composed primarily of DVPP nodes, can be coordinated to provide ancillary services such as FCR and virtual inertia support for limited periods (as part of N-1 contingency, for instance). For \textbf{S1}, the unmeasured active power at $\node{i\in\{1,2,3\}}$ varies in the following manner: $\unmeasuredpower{1}=0\xrightarrow[]{t=5s}0.015$ p.u., $\unmeasuredpower{2}=0\xrightarrow[]{t=15s}0.01$ p.u. and $\unmeasuredpower{3}=0\xrightarrow[]{t=25s}0.02$ p.u., while $\pielec$ and $\pirenewable$ are set to zero.  The power limits of BESS unit for each DVPP is defined as $\node{i}:\{\piibrmin,\piibrmax\} \ = \ \node{1}:\{0.02,-0.02\}, \ \node{2}:\{0.05,-0.05\}, \ \node{3}:\{0.01, -0.01\}$.  Plots for \textbf{S1} shown in Fig. \ref{fig:scenario1}, illustrate the capacity of DE  to accurately track the frequency and unmeasured active power while local resources compensate the mismatch to regulate frequency. The BESS power set-point $\idealcontrolsignal$ increases as $\hatpowerimbalancei$ becomes nonzero starting from $t=5 \ s$, while the $\distributedcontrolsignali$ at $\node{i=\{1,2\}}$ increase steadily in proportion starting @ $t=25 \ s$ due to a power deficit at $\node{3}$, with steady state values correponding to the set $\scalingfactordistributed{i}$.  The net tie-line power of node 4 increases momentarily  @ $t= 25 \ s$ to provide droop support before settling to zero as the neighbouring DVPPs ramp up their $\idealcontrolsignal$ to support node 3.  

The second scenario \textbf{S2}, considers random fluctuations in load and renewable power which is typical in power systems. The SG trips @ $t=10 \ s$ which causes a sudden decline in inertia at node 4 i.e. $\inertiaconstanti=4\xrightarrow[]{t=10s}0.005 \ s$ causing rapid fluctuations in the system frequency. We assume that the DVPPs are able to either automatically detect loss of generating unit and switch to additional inertia support or the network operator such as TSO detects the generator trip and invokes additional inertia support from the DVPPs.  The inertia constant for all the DVPP nodes increases as: $\inertiaconstant{i\in\{1,2,3\}}=0.01\xrightarrow[]{t=10s}0.1 \ s$. It is also assumed that the DVPP has information about its own inertia constant which is used directly in the DE.  The power limits on BESS unit and changes in $\unmeasuredpoweri$ remain the same as in \textbf{S1}.  Similar to \textbf{S1}, the distributed control action kicks in when $\actualcontrolsignal\neq\idealcontrolsignal$, either due to BESS dynamics (causing temporary jumps) or a shortfall in local resources (steady values). The same evaluation is repeated without additional inertia support from the DVPPs ($\inertiaconstanti=0.01 \ s$ throughout) and the average grid frequency is compared in the two cases. Clearly, the increased virtual inertia results in fewer rapid fluctuations compared to the second case.

\section{Conclusions and Outlook}
\label{section:conclusion}
This work introduces a two-layer coordinated fast frequency regulation scheme for DVPPs having local decentralised control enabled via an internal model based disturbance estimator and distributed compensation of local shortfall via DAPI approach. Simulation results illustrate the efficacy of the approach on a 4-bus system and performance improvement brought about by the inertia support enabled via DVPP, particularly in the event of contingencies. Investigating the impact of different estimator structures, such as those in \cite{ahmad2024TIA,ahmad2021active} and different consensus algorithms on frequency regulation can be an interesting outlook of this work. On the other hand, data-driven extensions can be explored to robustify against uncertain/unknown parameters.
\bibliographystyle{IEEEtran}
\bibliography{biblio}   

\begin{thebibliography}{10}
\providecommand{\url}[1]{#1}
\csname url@samestyle\endcsname
\providecommand{\newblock}{\relax}
\providecommand{\bibinfo}[2]{#2}
\providecommand{\BIBentrySTDinterwordspacing}{\spaceskip=0pt\relax}
\providecommand{\BIBentryALTinterwordstretchfactor}{4}
\providecommand{\BIBentryALTinterwordspacing}{\spaceskip=\fontdimen2\font plus
\BIBentryALTinterwordstretchfactor\fontdimen3\font minus \fontdimen4\font\relax}
\providecommand{\BIBforeignlanguage}[2]{{%
\expandafter\ifx\csname l@#1\endcsname\relax
\typeout{** WARNING: IEEEtran.bst: No hyphenation pattern has been}%
\typeout{** loaded for the language `#1'. Using the pattern for}%
\typeout{** the default language instead.}%
\else
\language=\csname l@#1\endcsname
\fi
#2}}
\providecommand{\BIBdecl}{\relax}
\BIBdecl

\bibitem{blackout2018australia}
R.~Yan, T.~K. Saha, F.~Bai, H.~Gu \emph{et~al.}, ``The anatomy of the 2016 south australia blackout: A catastrophic event in a high renewable network,'' \emph{IEEE Trans. Pow. Sys.}, vol.~33, no.~5, pp. 5374--5388, 2018.

\bibitem{justo2013ac}
J.~J. Justo, F.~Mwasilu, J.~Lee, and J.-W. Jung, ``Ac-microgrids versus dc-microgrids with distributed energy resources: A review,'' \emph{Renew. Sust. Ener. Rev.}, vol.~24, pp. 387--405, 2013.

\bibitem{naval2021virtual}
N.~Naval and J.~M. Yusta, ``Virtual power plant models and electricity markets-a review,'' \emph{Renew. Sust. Ener. Rev.}, vol. 149, p. 111393, 2021.

\bibitem{SeifVPPMarketsModelsOptChallengesOpportunities}
M.~M. Roozbehani, E.~Heydarian-Forushani, S.~Hasanzadeh, and S.~B. Elghali, ``Virtual power plant operational strategies: Models, markets, optimization, challenges, and opportunities,'' \emph{Sustainability}, vol.~14, no.~19, p. 12486, 2022.

\bibitem{dvpp2022florian}
B.~Marinescu, O.~Gomis-Bellmunt, F.~D{\"o}rfler, H.~Schulte, and L.~Sigrist, ``Dynamic virtual power plant: A new concept for grid integration of renewable energy sources,'' \emph{IEEE Access}, vol.~10, pp. 104\,980--104\,995, 2022.

\bibitem{lowInertiaChallenges2018}
F.~Milano, F.~D{\"o}rfler, G.~Hug, D.~J. Hill, and G.~Verbi{\v{c}}, ``Foundations and challenges of low-inertia systems,'' in \emph{2018 power systems computation conference (PSCC)}.\hskip 1em plus 0.5em minus 0.4em\relax IEEE, 2018, pp. 1--25.

\bibitem{dataDrivenFastFreqControlJWSporco}
E.~Ekomwenrenren, J.~W. Simpson-Porco, E.~Farantatos, M.~Patel, A.~Haddadi, and L.~Zhu, ``Data-driven fast frequency control using inverter-based resources,'' \emph{IEEE Transactions on Power Systems}, vol.~39, no.~4, pp. 5755--5768, 2023.

\bibitem{rlFreqContrl2018}
Z.~Yan and Y.~Xu, ``Data-driven load frequency control for stochastic power systems: A deep reinforcement learning method with continuous action search,'' \emph{IEEE Transactions on Power Systems}, vol.~34, no.~2, pp. 1653--1656, 2018.

\bibitem{rlFreqCtrl2022}
Z.~Yi, Y.~Xu, X.~Wang, W.~Gu, H.~Sun, Q.~Wu, and C.~Wu, ``An improved two-stage deep reinforcement learning approach for regulation service disaggregation in a virtual power plant,'' \emph{IEEE Transactions on Smart Grid}, vol.~13, no.~4, pp. 2844--2858, 2022.

\bibitem{mpc}
P.~R.~B. Monasterios and P.~Trodden, ``Low-complexity distributed predictive automatic generation control with guaranteed properties,'' \emph{IEEE Transactions on Smart Grid}, vol.~8, no.~6, pp. 3045--3054, 2017.

\bibitem{HiererchicalfastfreqControlJWSporco}
E.~Ekomwenrenren, Z.~Tang, J.~W. Simpson-Porco, E.~Farantatos, M.~Patel, and H.~Hooshyar, ``Hierarchical coordinated fast frequency control using inverter-based resources,'' \emph{IEEE Transactions on Power Systems}, vol.~36, no.~6, pp. 4992--5005, 2021.

\bibitem{swingEqnNature}
D.~Linaro, F.~Bizzarri, D.~Del~Giudice, C.~Pisani, G.~M. Giannuzzi, S.~Grillo, and A.~M. Brambilla, ``Continuous estimation of power system inertia using convolutional neural networks,'' \emph{Nature Communications}, vol.~14, no.~1, p. 4440, 2023.

\bibitem{ahmad2021active}
S.~Ahmad and A.~Ali, ``On active disturbance rejection control in presence of measurement noise,'' \emph{IEEE Trans. Ind. Electron.}, vol.~69, no.~11, pp. 11\,600--11\,610, 2021.

\bibitem{ahmad2023TIA}
S.~Ahmad and H.~Ahmed, ``Robust intrusion detection for resilience enhancement of industrial control systems: An extended state observer approach,'' \emph{IEEE Trans. Ind. Appl.}, vol.~59, no.~6, pp. 7735--7743, 2023.

\bibitem{ahmad2024TIA}
S.~Ahmad, R.~P. C.~d. Souza, P.~Kergus, Z.~Kader, and S.~Caux, ``Estimation-based robust switching control of a dc-dc boost converter,'' \emph{IEEE Trans. Ind. Appl.}, vol.~61, no.~1, pp. 1292--1304, 2025.

\bibitem{dapiInceptionJWSPorco2013Automatica}
J.~W. Simpson-Porco, F.~D{\"o}rfler, and F.~Bullo, ``Synchronization and power sharing for droop-controlled inverters in islanded microgrids,'' \emph{Automatica}, vol.~49, no.~9, pp. 2603--2611, 2013.

\bibitem{dapiStabilityJWSPorco2020IeeeCSS}
J.~W. Simpson-Porco, ``On stability of distributed-averaging proportional-integral frequency control in power systems,'' \emph{IEEE Control Systems Letters}, vol.~5, no.~2, pp. 677--682, 2020.

\end{thebibliography}
\end{document}